{\catcode`\@=11                                                  
   \gdef\SchlangeUnter#1#2{\lower2pt\vbox{\baselineskip 0pt\lineskip0pt    
   \ialign{$\m@th#1\hfil##\hfil$\crcr#2\crcr\sim\crcr}}}}           
\def\gtrsim{\mathrel{\mathpalette\SchlangeUnter>}}               
\def\lesssim{\mathrel{\mathpalette\SchlangeUnter<}}
\def\teq#1{$\, #1\,$}                           % text equation
\def\sigt{\hbox{$\sigma_{\hbox{\fiverm T}}$}}

\def\fsc{\alpha_{\hbox{\sevenrm f}}}                           
\def\erg{\varepsilon}
\def\LX{L_{\hbox{\sevenrm X}}}
\def\LXscal{L_{\hbox{\sevenrm X,35}}}
\def\effrad{\epsilon_{\hbox{\sevenrm rad}}}                           
\def\thetacap{\theta_{\hbox{\sevenrm cap}}}

\def\Gammacyc{\Gamma_{\hbox{\sevenrm cyc}}}   
\def\rhoGJ{\rho_{\hbox{\fiverm GJ}}}  
\def\Esd{{\dot E}_{\hbox{\fiverm SD}}}  
\def\dover#1#2{\hbox{${{\displaystyle#1 \vphantom{(} }\over{
   \displaystyle #2 \vphantom{(} }}$}}                
\font\fiverm=cmr5             \font\sevenrm=cmr7

\documentclass[twocolumn,runningheads]{svjour2_prep}
\smartqed  % flush right qed marks, e.g. at end of proof
   \usepackage{graphicx}
   \usepackage{epstopdf}
\journalname{Astrophysics and Space Science}
\begin{document}
\newcommand{\vol}[2]{$\,$\bf #1\rm , #2.}                 
\title{Resonant Compton Upscattering in Anomalous X-ray Pulsars
\thanks{This work was supported in part by the NASA INTEGRAL Theory 
program, and the NSF Stellar Astronomy and Astrophysics program 
through grant AST 0607651.}
}

\titlerunning{Hard X-rays from Anomalous X-ray Pulsars}        % if too long for running head

\author{Matthew G. Baring  \and Alice K. Harding}


\institute{Matthew G. Baring \at
              Rice University\\
              Department of Physics and Astronomy, MS 108\\
              6100 Main St.\\
              Houston, TX 77005, USA\\
%              Tel.: +1-713-348-2983\\
              \email{\it baring@rice.edu\rm}
           \and
              Alice K. Harding\at 
              Gravitational Astrophysics Laboratory\\
              Exploration of the Universe Division\\
              NASA Goddard Space Flight Center, Code 663\\
              Greenbelt, MD 20771, USA\\
%              Tel.: +1-301-286-7824\\
               \email{\it harding@twinkie.gsfc.nasa.gov\rm}
}

\date{October 12th, 2006} %Received: date / Accepted: date}

\maketitle

\begin{abstract}
A significant new development in the study of Anomalous X-ray Pulsars
(AXPs) has been the recent discovery by INTEGRAL and RXTE of flat, hard
X-ray components in three AXPs. These non-thermal spectral components
differ dramatically from the steeper quasi-power-law tails seen in the
classic X-ray band in these sources.  A prime candidate mechanism for
generating this new component is resonant, magnetic Compton
upscattering.  This process is very efficient in the strong magnetic
fields present in AXPs.  Here an introductory exploration of an inner
magnetospheric model for upscattering of surface thermal X-rays in AXPs
is offered, preparing the way for an investigation of whether such
resonant upscattering can explain the 20-150 keV spectra seen by
INTEGRAL.  Characteristically flat emission spectra produced by
non-thermal electrons injected in the emission region are computed using
collision integrals.   A relativistic QED scattering cross section is
employed so that Klein-Nishina reductions are influential in determining
the photon spectra and fluxes.  Spectral results depend strongly on the
magnetospheric locale of the scattering and the observer's orientation,
which couple directly to the angular distributions of photons sampled.
\end{abstract}

\section{Introduction}
 \label{intro}
Over the last decade, there has been a profound growth in evidence for a
new class of isolated neutron stars with ultra-strong magnetic fields,
so-called {\bf magnetars} that include Soft-Gamma Repeaters (SGRs) and
Anomalous X-ray Pulsars (AXPs).  Such a class was first postulated as a
model for SGRs by Duncan \& Thompson (1992), and later for AXPs
(Thompson \& Duncan 1996). The AXPs, are a group of six or seven
pulsating X-ray sources with periods around 6-12 seconds.  They are
bright, possessing luminosities \teq{\LX \sim 10^{35}\,\rm erg\;
s^{-1}}, show no sign of any companion, are steadily spinning down and
have ages \teq{\tau \lesssim 10^5} years (e.g. Vasisht \& Gotthelf
1997).  The steady X-ray emission has been clearly observed in a number
of AXPs (e.g. see Tiengo et al. 2002, for XMM observations of 1E
1048.1-5937; Juett et al. 2002 and Patel et al. 2003, for the {\it
Chandra} spectrum of 4U 0142+61), and also SGRs (see Kulkarni et al.
2003 for {\it Chandra} observations of the LMC repeater, SGR 0526-66). A
nice summary of spectral fitting of ASCA X-ray data from both varieties
of magnetars is given in Perna et al. (2001).  This emission displays
both thermal contributions, which have \teq{kT\sim 0.5-1} keV and so are
generally hotter than those in isolated pulsars, and also non-thermal
components with steep spectra that can be fit by power-laws
\teq{dn/dE\propto E^{-s}} of index in the range \teq{s=2-3.5}.

Flux variability in AXPs is generally small, suggesting that even the
non-thermal components experience a moderating influence of the stellar
surface, rather than some more dynamic dissipation in the larger
magnetosphere. Yet the recent observation (Gavriil \& Kaspi 2004) of
long-lived pulsed flux flares on the timescale of several months in AXP
1E 1048.1-5937 resembles earlier reports (Baykal \& Swank 1996;
Oosterbroek et al. 1998) of modest flux instability.  There are also
correlated long term variations in X-ray flux and non-thermal spectral
index in the source 1RXS J170849.0-400910, as identified by Rea et al.
(2005). Moreover, Kaspi et al. (2003) and Gavriil, Kaspi \& Woods (2002,
2004) reported bursting activity in the AXPs 1E 2259+586 and 1E
1048.1-5937, suggesting that anomalous X-ray pulsars are indeed very
similar to SGRs, a ``unification paradigm'' that is currently gathering
support, but remains to be established.

The recent detection by INTEGRAL and RXTE of hard, non-thermal pulsed
tails in three AXPs has provided an exciting new twist to the AXP
phenomenon.  In all of these, the differential spectra above 20 keV are
extremely flat: 1E 1841-045 (Kuiper, Hermsen \& Mendez 2004) has a
power-law energy index of \teq{s=0.94} between around 20 keV and 150
keV, 4U 0142+61  displays an index of \teq{s=0.2} in the 20 keV -- 50
keV band, with a steepening at higher energies implied by the total
DC+pulsed spectrum (Kuiper et al. 2006), and RXS J1708-4009 has
\teq{s=0.88} between 20 keV and 150 keV (Kuiper et al. 2006); these
spectra are all much flatter than the non-thermal spectra in the \teq{<
10} keV band.  Also, no clear tail has been seen in 1E 2259+586, yet
there is a suggestion of a turn-up in its spectrum in the interval
10--20 keV (Kuiper et al. 2006).  The identification of these hard tails
was enabled by the IBIS imager on INTEGRAL and secured by a review of
archival RXTE PCA and HEXTE data.  These tails do not continue much
beyond the IBIS energy window, since there are strongly constraining
upper bounds from Comptel observations of these sources that necessitate
a break and steepening somewhere in the 150--750 keV band (see Figures
4, 7 and 10 of Kuiper et al. 2006).  Interestingly, Molkov et al. (2004)
and Mereghetti et al. (2004) also reported evidence for hard tails in
SGR 1806-20, so that the considerations here are germane also to SGRs in
quiescence.

Explaining the generation of these hard tails forms the motivation for
this paper, which presents an initial exploration of the production of
non-thermal X-rays by inverse Compton heating of soft, atmospheric
thermal photons by relativistic electrons. The electrons are presumed to
be accelerated either along open or closed field lines, perhaps by
electrodynamic potentials, or large scale currents associated with
twists in the magnetic field structure (e.g. see Thompson \& Beloborodov
2005). In order to power the AXP emission, they must be produced with
highly super-Goldreich-Julian densities.  In the strong fields of the
inner magnetospheres (i.e. within 10 stellar radii) of AXPs, the inverse
Compton scattering is predominantly resonant at the cyclotron frequency,
with an effective cross section above the classical Thomson value. 
Hence, proximate to the neutron star surface, in regions bathed
intensely by the surface soft X-rays, this process can be extremely
efficient for an array of magnetic colatitudes.  Here, an investigation
of the general character of emission spectra is presented, using
collision integral analyses that will set the scene for future
explorations using Monte Carlo simulations.  This scenario forms an
alternative to recent proposals (Thompson \& Beloborodov 2005; Heyl \&
Hernquist 2005) that the new components are of synchrotron or
bremsstrahlung origin, and at higher altitudes than considered here. 
The efficiency of the resonant Compton process suggests it will dominate
these other mechanisms if the site of electron acceleration is
sufficiently near the stellar surface.  This prospect motivates the
investigation of resonant inverse Compton models.

\section{The Compton Resonasphere}
 \label{sec:resonasphere}

\subsection{Energetics}
 \label{sec:energetics}

The scattering scenario for AXP hard X-ray tail formation investigated
here assumes that the seed energization of electrons arises within a few
stellar radii of the magnetar surface.  This can in principal occur on
either open or closed field lines, so both possibilities will be
entertained.  The key requirement is the presence of ultra-relativistic
electrons moving along {\bf B}, with an abundance satisfying the
energetics of AXPs implied by their intense X-ray luminosities, \teq{\LX
\gtrsim 10^{35}}erg/sec above 10 keV (Kuiper et al. 2006).  The hard
X-ray tails have luminosities that are 2--3 orders of magnitude greater
than the classical spin-down luminosity  \teq{\Esd\sim 8\pi^4
B_p^2R^6/(3P^4c^3)}  due to magnetic dipole radiation
torques.  Here \teq{B_p} is the surface polar field strength.  
This signature indicates that other dissipation mechanisms,
such as structural rearrangements of crustal magnetic fields, power the
AXP emission (e.g. Thompson \& Duncan 1995; 1996).  Let \teq{n_e} be the
number density of such electrons, \teq{\langle \gamma_e\rangle} be their
mean Lorentz factor, and \teq{\effrad} be the efficiency of them
radiating during their traversal of the magnetosphere (either along open
or closed field lines). Then one requires that \teq{\LX \sim \effrad
\langle \gamma_e\rangle m_ec^2 (4\pi n_e R_c^2c)} if the emission column
has a base that is a spherical cap of radius \teq{R_c}.  This yields
number densities \teq{n_e\sim 3\times 10^{17} \LXscal /\effrad \langle
\gamma_e\rangle} cm$^{-3}$ for scaled luminosities \teq{\LXscal \equiv
\LX /10^{35}}erg/sec, if \teq{R_c\sim 10^6}cm. Therefore large densities
are needed, though not impossible ones, since optically thin conditions
for the surface thermal X-rays prevail provided that \teq{\langle
\gamma_e\rangle\gg 1}, and \teq{\effrad} is not miniscule.

Comparing \teq{en_e} to the classic Goldreich-Julian (1969) density
\teq{\rhoGJ =\nabla .\vec{E}/4\pi = -\vec{\Omega}.\vec{B}/(2\pi c)} for
force-free, magnetohydrodynamic rotators, one arrives at the ratio 
\begin{equation}
   \dover{en_e}{\vert\rhoGJ\vert} \;\approx\; \dover{4,670}{\effrad \langle \gamma_e\rangle}\;
   \dover{L_{X,35}\, P}{B_{15} R_6^2}\quad ,
 \label{eq:GJ_compare}
\end{equation}
for AXP pulse periods \teq{P} in units of seconds, polar magnetic fields
\teq{B_{15}} in units of \teq{10^{15}}Gauss, and cap radii \teq{R_6} in
units of \teq{10^6}cm.  Large electron Lorentz factors of
\teq{\gamma_e\gg 10^2-10^3} and their efficient resonant Compton cooling
(i.e. \teq{\effrad\sim 0.01 - 1}) are readily attained in isolated
pulsars with \teq{B\sim 0.1} (e.g. see Sturner 1995; Harding \& Muslimov
1998; Dyks \& Rudak 2000), and such conditions are expected to persist
into the magnetar field regime. For \teq{\effrad \langle
\gamma_e\rangle\gtrsim 10^3} and \teq{R_6\sim 1}, the requisite density
\teq{n_e} is super-Goldreich-Julian, but not dramatically so.  This
situation corresponds, however, to acceleration zone cap radii \teq{R_c}
that are considerably larger than standard polar cap radii
\teq{R\thetacap} for AXPs.   Here \teq{\thetacap = \arcsin \{(2\pi
R/Pc)^{1/2}\} } for a pulsar of radius \teq{R} and period \teq{P}.  For
\teq{P=10}sec, this yields \teq{R\thetacap\sim 4.6\times 10^3}cm. 
Concentrating the relativistic electrons in such a narrow column yields
charge densities far exceeding the Goldreich-Julian benchmark implied by
Eq.~(\ref{eq:GJ_compare}).  However, since AXPs possess luminosities 
\teq{\LX \gg \Esd}, profound collatitudinal confinement may not prove 
necessary.  Energetically, non-dipolar structure at the surface is
easily envisaged, since the dissipation mechanism that powers AXP
emission can restructure the fields; there is suggestive evidence for
these reconfigurations from variations seen in the pulse profiles,
particularly after flaring activity (e.g. see Kaspi et al. 2003).  The
electron energization zone may then cover a much larger range of
colatitudes than is assigned to a standard polar cap, and may extend to
closed field lines in equatorial regions. Of course, deviations from
dipole structure will modify the contours for the resonasphere
considerably from those illustrated just below, though mostly in the
inner magnetosphere.

\subsection{Resonant Compton Scattering}
 \label{sec:rescomp}

In strong neutron star fields the cross section for Compton scattering
is resonant at the cyclotron energy and a series of higher harmonics
(e.g. see Daugherty \& Harding 1986), effectively increasing the
magnitude of the process over the Thomson cross section \teq{\sigt} by
as much as the order of \teq{1/(\fsc B)}, where \teq{\fsc} is the fine
structure constant.   Here, as throughout the paper, magnetic fields are
written in units of \teq{B_{\rm cr}=m_e^2c^3/(e\hbar )=4.413\times
10^{13}}Gauss, the quantum critical field strength.  Klein-Nishina-like
declines operate in supercritical fields, reducing the effective cross
section at the resonance (e.g. Gonthier et al. 2000).  In the
non-relativistic, Thomson regime (e.g. see Herold 1979), only the
fundamental resonance is retained.  For the specific case of
ultra-relativistic electrons colliding with thermal X-rays, in the
electron rest frame (ERF), the photons initially move mostly almost
along {\bf B}, and the cyclotron fundamental is again the only resonance
that contributes (Gonthier et al. 2000).

The dominance of this resonance in forming upscattering spectra leads to
an effective kinematic coupling between the energies \teq{\erg_{\gamma}
m_ec^2}  and \teq{\gamma_e m_ec^2} of colliding photons and electrons,
respectively, and the angle of the initial photon \teq{\theta_{\gamma}}
to the magnetic field lines: the cyclotron fundamental is sampled when 
\begin{equation}
   \gamma_e\erg_{\gamma} (1-\cos\theta_{\gamma}) \;\approx\; B\quad , \quad 
   \hbox{for}\; \gamma_e\;\gg\; 1\quad .
 \label{eq:res_kinematics}
\end{equation}
The simplicity of this coupling automatically implies that integration
over an angular distribution of incoming photons results in a
flat-topped emission spectrum for Compton upscattering of isotropic
photons in strong magnetic fields.  This characteristic is
well-documented in the literature in the magnetic Thomson limit (e.g.
see Dermer 1990; Baring 1994; for old gamma-ray burst scenarios, and
Daugherty \& Harding 1989; Sturner, Dermer \& Michel 1995 for pulsar
contexts), specifically for collisions between \teq{\gamma_e\gg 1}
electrons and thermal X-rays emanating from a neutron star surface.

\begin{figure}
   \centering
   \includegraphics[width=0.48\textwidth]{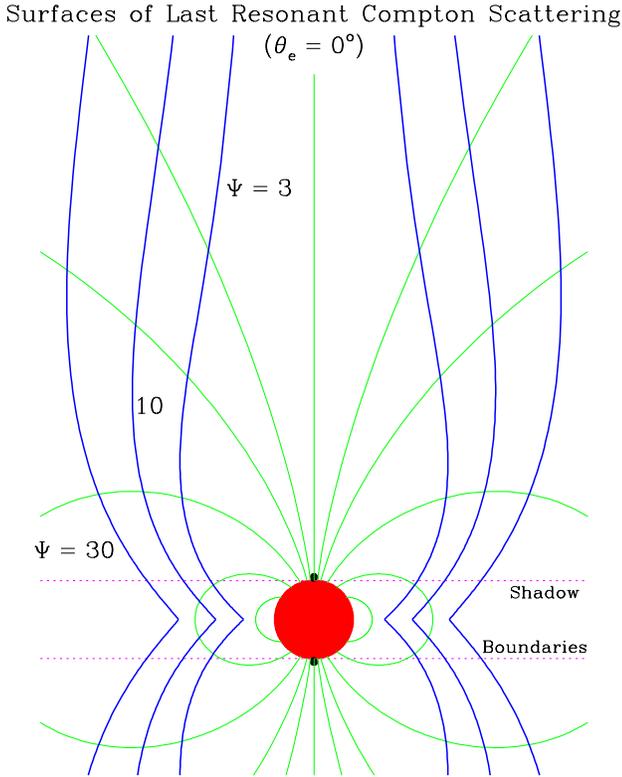}
   \caption{Contours in a section of a pulsar magnetosphere that
   depict cross sections of the surfaces of last resonant scattering,
   i.e. the maximal extent of the Compton resonasphere.  The heavyweight
   contours (in dark blue) are computed for different values of the 
   resonance parameter \teq{\Psi} defined in Eq.~(\ref{eq:surface_locus}), 
   and at extremely high altitudes asymptotically approach the magnetic
   axis (vertical line). The filled red circle denotes the neutron star,
   whose radius \teq{R} establishes the spatial scale for the figure.
   The case illustrated is for photons emanating from the polar axis
   (i.e. \teq{\theta_e=0^{\circ}}), denoted by black dots, for which the
   neutron star shadow regions are demarcated by the dotted boundaries,
   and only the surfaces (azimuthally-symmetric about the magnetic axis)
   are accessible to resonant Compton interactions. In computing the
   contours, the upscattering electrons were assumed ultra-relativistic,
   and the depicted spatial scales are linear.
   }
 \label{fig:resonasphere} 
\end{figure}

It is instructive to compute the zones of influence of the resonant
Compton process for magnetic dipole field geometry.  For X-ray photons
with momentum vector \teq{\vec{k}}, emanating from a single point at
position vector \teq{\vec{R}_e} on the stellar surface with colatitude
\teq{\theta_e} (i.e., \teq{\theta_e=0} corresponds to the magnetic
pole), the kinematic criterion in Eq.~(\ref{eq:res_kinematics}) selects
out a single photon angle \teq{\theta_{\gamma}} for a given local
\teq{B}, both of which are dependent on the altitude and colatitude of
the point of interaction.  This assumes that the photon propagates with
no azimuthal component to its momentum, a specialization that will be
remarked upon shortly.  In the absence of rotation, the resonance
criterion defines a surface that is azimuthally symmetric about the
magnetic field axis. The locus of the projection of this surface onto a 
plane intersecting the magnetic axis can be found through elementary 
geometry, assuming a flat spacetime.  Observe that light bending due to 
the stellar gravitational potential will modify these loci significantly in
inner equatorial regions. At an altitude \teq{r} and colatitude
\teq{\theta}, denoted by position vector \teq{\vec{r}}, the magnetic
dipole polar coordinate components are
\begin{equation}
   B_r\; =\; \dover{B_pR^3}{r^3}\, \cos\theta\quad ,\quad
   B_{\theta} \; =\; \dover{B_pR^3}{2r^3}\, \sin\theta\quad ,
 \label{eq:dipole_field}
\end{equation}
where \teq{R} is the neutron star radius and \teq{B_p} is the surface field 
strength at the magnetic pole.  At this location \teq{\vec{r}}, the corresponding 
polar coordinate components \teq{k_r} and \teq{k_{\theta}} of the photon 
momentum are given by
\begin{eqnarray}
  \dover{k_r}{\vert\vec{k}\vert} &=& \dover{\chi -\cos (\theta -\theta_e)}{
       \sqrt{1-2\chi\cos (\theta -\theta_e) +\chi^2}}\quad , \nonumber\\[-5.5pt]
 \label{eq:kvec_components}\\[-5.5pt]
  \dover{k_{\theta}}{\vert\vec{k}\vert} &=& \dover{\sin (\theta -\theta_e)}{
       \sqrt{1-2\chi\cos (\theta -\theta_e) +\chi^2}}\quad , \nonumber
\end{eqnarray}
where \teq{\chi=r/R} is the scaled altitude. The geometry of the
magnetic dipole then uniquely determines the angle
\teq{\theta_{\gamma}=\theta_{\gamma}(r/R,\, \theta ;\, \theta_e)} of the
photon to the field via the relation \teq{\cos\theta_{\gamma} = \vec{k}
. \vec{B} / \vert \vec{k} \vert . \vert \vec{B} \vert =
(k_rB_r+k_{\theta}B_{\theta})/ \vert \vec{k} \vert . \vert \vec{B}
\vert}:
\begin{equation}
   \cos\theta_{\gamma}\; =\; \dover{2 \cos\theta [\chi -\cos (\theta -\theta_e)]
       + \sin\theta\,\sin (\theta -\theta_e)}{
   \sqrt{1+3\cos^2\theta}\; \sqrt{1-2\chi\cos (\theta -\theta_e) +\chi^2}}\; .
 \label{eq:theta_gamma}
\end{equation}
Inserting this into Eq.~(\ref{eq:res_kinematics}), for the specific case
of \teq{\theta_e=0}, yields the equation 
for the locus defining the surface of resonant scattering for 
outward-going electrons: 
\begin{equation}
   \chi^3\; =\; \Psi\; \dover{\sqrt{1+3\cos^2\theta}}{
       1-\cos\theta_{\gamma}}\quad , \quad
    \Psi\; =\; \dover{B_p}{2\gamma_e\erg_{\gamma}}\quad .
 \label{eq:surface_locus}
\end{equation}
Here \teq{\Psi} is the key parameter that scales the altitude of the
resonance locale, and typically might be in the range \teq{1-10^3} for
magnetars of \teq{\gamma_e\sim 10^2 - 10^4}.
Eq.~(\ref{eq:surface_locus}) can be rearranged into polynomial form, but
must be solved numerically.  For the special case of soft photons
emitted from the surface pole (\teq{\theta_e=0}), the surfaces of
resonant scattering for different \teq{\Psi} are illustrated in
Fig.~\ref{fig:resonasphere} as the heavy blue contours.  The shadows of
the emission point are also indicated to demarcate the propagation
exclusion zone for the chosen emission colatitude.

The altitude of resonance is clearly much lower in the equatorial
regions, since the photons tend to travel more across field lines in the
observer's frame, and so access the resonance in regions of higher field
strength.  In contrast, at small colatitudes above the magnetic pole,
\teq{\theta_{\gamma}} is necessarily small, pushing the resonant surface
to very high altitudes where the field is much lower.  In this case, the
loci asymptotically approach the polar axis at infinity, satisfying
\teq{\chi^3\sin^2\theta \approx 16\Psi} for the depicted case of
\teq{\theta_e=0}. For \teq{\theta_e \neq 0} cases (not depicted), the
contours are morphologically similar, though they incur significant
deviations from those in Fig.~\ref{fig:resonasphere}, both in equatorial
and polar regions; for example, when \teq{\theta\to 0}, the loci do not
extend to infinity for \teq{\theta_e > 0}. Clearly, by sampling
different emission colatitudes \teq{\theta_e} these surfaces are smeared
out into annular volumes.  Furthermore, since most photons not emitted
at the poles possess azimuthal components to their momenta, propagation
out of the plane of the diagram must also be considered, modifying
Eqs.~(\ref{eq:theta_gamma}) and~(\ref{eq:surface_locus}).  In the
interests of compactness, such algebra is not offered here.  It suffices
to observe that introducing an azimuthal component to the photon
momentum generally tends to increase propagation across the field in the
observer's frame, i.e. increasing \teq{\theta_{\gamma}}, so that the
resonance criterion in Eq.~(\ref{eq:res_kinematics}) is realized at
lower altitudes and higher field locales.  Hence, taking into account
azumithal contributions to photon propagation in the magnetosphere, loci
like those depicted in Fig.~\ref{fig:resonasphere} actually represent
the outermost extent of resonant interaction, and so are {\it surfaces
of last resonant scattering}, i.e. the outer boundaries to the {\bf
Compton resonasphere}.  It is evident that, for the majority of closed
field lines for long period AXPs, this resonasphere is confined to
within a few stellar radii of the surface. Clearly, introducing more
complicated, non-dipolar field topologies will also tend to lower the
altitudes of the resonasphere.

\section{Compton Upscattering Spectra in AXPs}
 \label{sec:upscattering}

To gain an initial idea of what emission spectra might be produced in
the resonasphere of AXPs, collision integral calculations of
upscattering spectra are performed. Here results are presented for
monoenergetic electrons of Lorentz factor \teq{\gamma_e}, from which
spectral forms for various electron distributions can easily be
inferred.  This approach forgoes considerations of electron cooling,
which naturally generates quasi-power-law distributions, at least in the
magnetic Thomson limit (e.g. see Dermer 1990; Baring 1994); such issues
will be addressed for supercritical fields in future presentations. To
simplify the formalism for the spectra, monoenergetic incident photons
of dimensionless energy \teq{\erg_{\gamma}=\erg_s} will be assumed, with
the implicit understanding that \teq{\erg_s\sim 3kT/m_ec^2}, so that
values of \teq{\erg_s\sim 0.003} are commensurate with thermal photon
temperatures \teq{kT\sim 0.5-1}keV observed in AXPs (see Perna et al.
2001). Distributing \teq{\erg_s} via a Planck spectrum provides only
small changes to the spectra illustrated here, serving only to smear out
modest spectral structure at the uppermost emergent photon energies.

Central to the characteristics of the spectral shape for resonant
upscattering problems are the kinematics associated with both the
Lorentz transformation from the observer's or laboratory frame (OF) to
the electron rest frame (ERF), and the scattering kinematics in the ERF.
 Since choices of photon angles in the two reference frames are not
unique, a statement of the conventions adopted here is now made to
remove any ambiguities.  Let the electron velocity vector in the OF be 
\teq{\vec{\beta}_e}. This will be parallel to {\bf B} due to rampant
cyclo-synchrotron cooling perpendicular to the field.  The dimensionless
pre- and post-scattering photon energies (i.e. scaled by \teq{m_ec^2})
in the OF are \teq{\erg_i} and \teq{\erg_f}, respectively, and the
corresponding angles of these photons with respect to
\teq{-\vec{\beta}_e} (i.e. field direction) are \teq{\Theta_i} and
\teq{\Theta_f}, respectively.  Observe that \teq{\Theta_i\to
\theta_{\gamma}} establishes a connection to the notation used in
Section~\ref{sec:resonasphere}.  With this definition,
\begin{equation}
   \cos\Theta_{i,f}\; =\; - \dover{\vec{\beta}_e . \vec{k}_{i,f}}{
        \vert\vec{\beta}_e\vert . \vert \vec{k}_{i,f}\vert }\quad ,
 \label{eq:OF_kinematics}
\end{equation}
and the zero angles are chosen anti-parallel to the electron velocity.
Here, \teq{\vec{k}_i} and \teq{\vec{k}_f} are the initial and final
photon three-momenta in the OF. Boosting by \teq{\vec{\beta}_e} to the
ERF then yields pre- and post-scattering photon energies in the ERF of
\teq{\omega_i} and \teq{\omega_f}, respectively, with corresponding
angles with respect to \teq{-\vec{\beta}_e} of \teq{\theta_i} and
\teq{\theta_f}.  The relations governing this Lorentz transformation are
\begin{eqnarray}
   && \omega_{i,f} \; =\; \gamma_e\erg_{i,f} (1+\beta_e\cos\Theta_{i,f})
   \quad ,\nonumber\\[-5.5pt]
 \label{eq:Lorentz_transform}\\[-5.5pt]
   && \cos\theta_{i,f} \; =\; \dover{\cos\Theta_{i,f} + \beta_e}{
      1 + \beta_e \cos\Theta_{i,f}}\quad .\nonumber
\end{eqnarray}
The inverse transformation relations are obtained from these by the
interchange \teq{\theta_{i,f}\leftrightarrow\Theta_{i,f}} and the
substitutions \teq{\omega_{i,f}\to\erg_{i,f}} and \teq{\beta_e\to
-\beta_e}.  The form of Eq.~(\ref{eq:Lorentz_transform}) guarantees that
for most \teq{\Theta_i}, the initial scattering angle \teq{\theta_i} in
the ERF is close to zero when \teq{\gamma_e\gg 1}, exceptions being
cases when \teq{\cos\Theta_i\approx -\beta_e}. These exceptional cases
form a small minority of the upscattering phase space, and indeed a
small contribution to the emergent spectra, and so are safely neglected
in the ensuing computations. This \teq{\theta_i\approx 0} approximation
yields dramatic simplification of the differential cross section for
resonant Compton scattering, and motivates the particular laboratory
frame angle convention adopted in Eq.~(\ref{eq:OF_kinematics}).

The scattering kinematics in the ERF differ from that described by the
familiar Compton formula in the absence of magnetic fields (e.g. see
Herold 1979; Daugherty \& Harding 1986).  In the special case
\teq{\theta_i\approx 0} that is generally operable for the scattering
scenario here, the kinematic formula for the final photon energy
\teq{\omega_f} in the ERF can be approximated by
\begin{equation}
   \omega_f\; =\; \omega' (\omega_i ,\,\theta_f)\;\equiv\;
     \dover{2\omega_i\, r}{1+\sqrt{1-2\omega_i r^2\sin^2\theta_f}}\quad ,
 \label{eq:reson_kinematics}
\end{equation}
where
\begin{equation}
   r\; =\; \dover{1}{1+\omega (1-\cos\theta_f)}
 \label{eq:Compton_formula}
\end{equation}
is the ratio \teq{\omega_f/\omega_i} that would correspond to the
non-magnetic Compton formula, which in fact does result if \teq{\omega_i
r^2\sin^2\theta_f\ll 1}.  Eq.~(\ref{eq:reson_kinematics}) can be found
in Eq. ~(15) of Gonthier et al. (2000), and is realized for the
particular case where electrons remain in the ground state (zeroth
Landau level) after scattering.  Such a situation occurs for the
resonant problem addressed in this paper, a feature that is discussed
briefly below.

Let \teq{n_{\gamma}} be the number density of photons resulting from the
resonant scattering process. For inverse Compton scattering, an
expression for the spectrum of photon production \teq{dn_{\gamma}/(dt\,
d\erg_f\, d\mu_f) }, differential in the photon's post-scattering
laboratory frame quantities \teq{\erg_f} and \teq{\mu_f=\cos\Theta_f},
was presented in Eqs.~(A7)--(A9) of Ho and Epstein (1989), valid for
general scattering scenarios, including Klein-Nishina regimes.  This was
used by Dermer (1990) and Baring (1994) in the magnetic Thomson domain,
i.e. when \teq{B\ll 1} and the photon energy in the electron rest frame
(ERF) is far inferior to \teq{m_ec^2}.   Such specialization is readily
extended to the magnetar regime by incorporating into the Ho and Epstein
formalism the magnetic kinematics and the QED cross section for fully
relativistic cases. The result can be integrated over \teq{\mu_f} and
then written as
\begin{eqnarray}
   \dover{dn_{\gamma}}{dt\, d\erg_f} &=&  \dover{n_e n_s\, c}{\mu_+-\mu_-}
   \int_{\mu_l}^{\mu_u}d\mu_f   \int_{\mu_-}^{\mu_+}d\mu_i \nonumber\\[-5.5pt]
 \label{eq:scatt_spec}\\[-5.5pt]
   &&\delta \big\lbrack\omega_f -\omega'(\omega_i,\,\theta_f)\, \bigl\rbrack\,
   \dover{1+\beta_e\mu_i}{\gamma_e(1+\beta_e\mu_f)}\,
   \dover{d\sigma}{d(\cos\theta_f) }\quad ,\nonumber
\end{eqnarray}
noting that the angle convention specified in
Eq.~(\ref{eq:OF_kinematics}) requires the substitution \teq{\beta_e\to
-\beta_e} in Eqs.~(A7--A9) of Ho and Epstein (1989). Here, the notation
\teq{\mu_i=\cos\Theta_i} and \teq{\mu_f=\cos\Theta_f} is used for
compactness, \teq{n_e} is the number density of relativistic electrons,
and \teq{n_s} is that for the soft photons.   These incident
monoenergetic photons are assumed to possess a uniform distribution of
angle cosines \teq{\mu_i} in some range \teq{\mu_-\leq\mu_i\leq \mu_+},
which is generally broad enough to encompass the resonance, i.e. the
value \teq{\mu_i=[ B/(\gamma_e\erg_s) -1]/\beta_e}. The bounds on the
\teq{\mu_f} integration, defining the observable range of angle cosines
with respect to the field direction, will be specialized to
\teq{\mu_l=-1} and \teq{\mu_u=1} in the illustration here, though
dependence of the emergent spectra \teq{\Theta_f} values will be
discussed below. The function \teq{\omega'(\omega_i,\,\theta_f)}, which
appears in the delta function in Eq.~(\ref{eq:scatt_spec}) that
encapsulates the scattering kinematics, is that defined in
Eq.~(\ref{eq:reson_kinematics}).

Fully relativistic, quantum cross section formalism for the Compton
interaction in magnetic fields can be found in Herold (1979), Daugherty
\& Harding (1986), and Bussard, Alexander \& M\'esz\'aros (1986). These
extend earlier non-relativistic quantum mechanical formulations such as
in Canuto, Lodenquai \& Ruderman (1971), and Blandford \& Scharlemann
(1976).  The differential cross section, \teq{d\sigma/d \cos \theta_f},
appearing in Eq.~(\ref{eq:scatt_spec}) is taken from Eq.~(23) of
Gonthier et al. (2000), and incorporates the relativistic QED physics. 
Yet, it is specialized to the case of scatterings that leave the
electron in the ground state, the zeroth Landau level that it originates
from.  This expedient choice is entirely appropriate, since it yields
the dominant contribution to the cross section at and below the
cyclotron resonance (e.g. Daugherty \& Harding 1986; Gonthier et al.
2000).  For considerations below, where information on the final
polarization state of the photon is retained, Eq.~(22) of Gonthier et
al. (2000) is used; this simplifies to the forms
\begin{eqnarray}
   \dover{d\sigma_{\parallel,\perp }}{d\cos \theta_f} & = & 
   \dover{3\sigma_{\rm T}}{32} \, 
   \dover{ (\omega_f)^3\; T_{\parallel,\perp }\,
                  \exp \bigl\{ -\omega_f^2\sin^2\theta_f/[2B] \bigr\} }{ \omega_i\, 
        [1+\omega_i (1-\cos \theta_f) - \omega_f\sin ^2\theta_f ]}\nonumber\\[-5.5pt]
  \label{eq:sigma_pol}\\[-5.5pt]
   & \times & \biggl\{ \dover{1}{(\omega_i -B)^2 + (\Gammacyc/2)^2} + 
                \dover{1}{(\omega_i +B - \zeta)^2} \biggr\}  \nonumber
\end{eqnarray}
for the differential cross sections, where the cyclotron decay width
\teq{\Gammacyc} (discussed below) has been introduced to render the 
resonance finite in the form of a Lorentz profile.  The exponential factor
in Eq.~(\ref{eq:sigma_pol}) is a relic of the Laguerre functions that
signal the discretization of momentum/energy states perpendicular to
the field.  Furthermore, the factor in square brackets in the denominator
is always positive, being proportional to \teq{1-\omega_f r\sin^2\theta_f},
which can be shown to be precisely the square root appearing in 
Eq.~(\ref{eq:reson_kinematics}).  Also, \teq{\zeta = \omega_i\omega_f 
(1-\cos\theta_f)}, and 
\begin{eqnarray}
   T_{\parallel} & = & 2\cos^2\theta_f+ \omega_i (1-\cos\theta_f)^2
                                                      -\omega_f\sin^2\theta_f\nonumber\\[-5.5pt]
 \label{eq:TparTperp}\\[-5.5pt]
   T_{\perp} & = & 2 + \omega_i (1-\cos\theta_f)^2
                                                 -\omega_f\sin^2\theta_f\quad .\nonumber
\end{eqnarray}
Here, the standard convention for the labelling of the photon linear
polarizations is adopted: \teq{\parallel} refers to the state with the
photon's {\it electric} field vector parallel to the plane containing
the magnetic field and the photon's momentum vector, while \teq{\perp}
denotes the photon's electric field vector being normal to this plane.

For magnetic Compton scattering, in the particular case of photons
propagating along {\bf B} prior to scattering (i.e. \teq{\theta_i =0}),
the differential cross sections are independent of the initial
polarization of the photon (Gonthier et al. 2000); this property is a
consequence of circular polarizations forming the natural basis states
for \teq{\theta_i =0}. Therefore, transitions \teq{\perp\to\parallel}
and \teq{\parallel\to\parallel} yield identical forms for the cross
sections, and separately so do the transitions \teq{\perp\to\perp} and
\teq{\parallel\to\perp}. Accordingly, the cross sections in
Eq.~(\ref{eq:sigma_pol}) are labelled only by the post-scattering linear
polarization state of the photon; they are summed when
polarization-independent results are desired, i.e. \teq{d\sigma/d \cos
\theta_f=(d\sigma_{\parallel}/d \cos \theta_f +d\sigma_{\perp}/d \cos
\theta_f)}.  Therefore, clearly the upscattered photon spectra presented
here are insensitive to the initial polarization level (zero or
otherwise) of the soft photons.

The integrals in Eq.~(\ref{eq:scatt_spec}) can be manipulated using the
Jacobian identity \teq{d\mu_i\, d\mu_f =d\omega_i\,
d\omega_f/(\gamma_e^2\beta_e^2\erg_i\erg_f)} to change variables to
\teq{\omega_i} and \teq{\omega_f}.  Observe that the values of
\teq{\mu_{\pm}} do not impact these integrations provided that the
resonance condition in Eq.~(\ref{eq:res_kinematics}) is sampled. The
\teq{\omega_f} integration is then trivial.  The \teq{\omega_i}
integration is more involved, but can be developed by suitable
approximation, as follows. The relativistic Compton cross section is
strongly peaked at the cyclotron fundamental (see Fig.~2 of Gonthier et
al. 2000) due to the appearance of the resonant denominator
\teq{1/[(\omega_i-B)^2 + (\Gammacyc/2)^2]}, where \teq{\Gammacyc\ll B}
is the dimensionless cyclotron decay rate from the first Landau level. 
Therefore, this Lorentz profile can be approximated by a delta function
in \teq{\omega_i} space of identical normalization in an integration
over \teq{\omega_i}:
\begin{equation}
   \dover{1}{(\omega_i -B)^2 + (\Gammacyc/2)^2}\;\to\;
   \dover{2\pi}{\Gammacyc}\, \delta (\omega_i-B)\quad .
 \label{eq:resonance_approx}
\end{equation}
This mapping, which was adopted in the magnetic Thomson limit by Dermer
(1990), renders the \teq{\omega_i} trivial, with the non-resonant term
in the cross section in Eq.~(\ref{eq:sigma_pol}) being neglected, and
the evaluation of the integrals in Eq.~(\ref{eq:scatt_spec}) complete. 
The spectra scale as the inverse of the decay rate \teq{\Gammacyc},
whose form can be found, for example, in Eqs.~(13) or (23) of Baring,
Gonthier \& Harding (2005; see also Latal 1986; Harding \& Lai 2006). 
For \teq{B\ll 1}, \teq{\Gammacyc\approx 4\fsc B^2/3}, which traces
classical cyclotron cooling, while for \teq{B\gg 1}, quantum effects and
recoil reductions generate \teq{\Gammacyc\approx (\fsc/e) \sqrt{B/2}}.
\begin{figure}[h]
   \vspace{-7pt}
   \centering
   \includegraphics[width=0.47\textwidth]{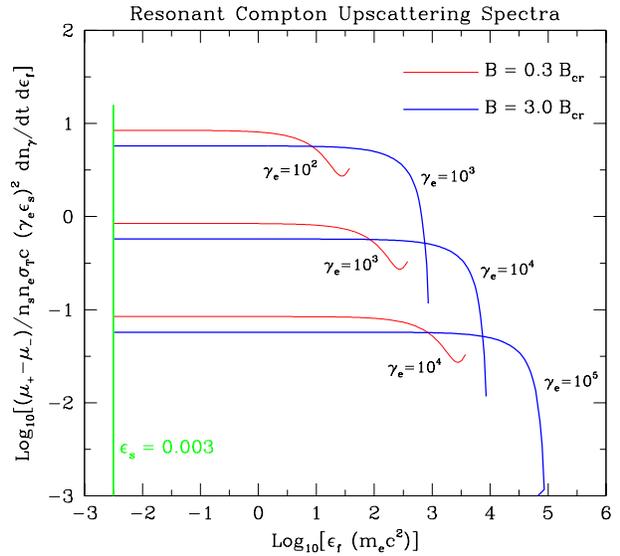}
   \vspace{-8pt}
   \caption{Resonant Compton upscattering spectra (scaled) such as 
   might be sampled in the magnetosphere of an AXP, for different 
   relativistic electron Lorentz factors \teq{\gamma_e}, as labelled. 
   The emergent photon energy \teq{\erg_f} is scaled in terms of 
   \teq{m_ec^2}.  The chosen magnetic field strengths of \teq{B=3 B_{\rm cr}} 
   (heavyweight, blue) and \teq{B=0.3 B_{\rm cr}} (lighter weight, red) 
   correspond to different altitudes and perhaps colatitudes.  
   Results are depicted for seed photons of energy \teq{\erg_s=0.003}
   (marked by the green vertical line), typical of thermal X-rays emanating 
   from AXP surfaces; downscattering resonant emission at 
   \teq{\erg_f < \erg_s} was not exhibited. 
   }
 \label{fig:upscatt_spectra} 
\end{figure}

Representative spectral forms are depicted in
Fig.~\ref{fig:upscatt_spectra}, for the situation where the emergent
polarization is not observed. Because of the approximation to the
resonance in Eq.~(\ref{eq:resonance_approx}), non-resonant scattering
contributions were omitted when generating the curves; these
contributions produce steep wings to the spectra at the uppermost and
lowermost energies (not shown), and a slight bolstering of the resonant
portion.  This resonant restriction suffices for the purposes of this
paper, and kinematically limits the range of emergent photon energies
\teq{\erg_f} to
\begin{equation}
   \gamma_e(1-\beta_e)\, B \;\leq\;\erg_f\;\leq\; 
   \dover{\gamma_e(1+\beta_e)\, B}{1+2B} \quad .
 \label{eq:kinematic_range}
\end{equation}
This range generally extends below the thermal photon seed energy
\teq{\erg_s}. Such downscatterings correspond to forward scatterings
\teq{\cos\theta_f\approx 1} in the ERF such that \teq{\theta_f\lesssim
\theta_i\sim 1/\gamma_e}.  Since they constitute a miniscule portion of
the angular phase space (and energy budget), only upscattering spectra
are exhibited in the Figure.

In the \teq{B=0.3} case, a quasi-Thomson regime, the spectra show a
characteristic flat distribution that is indicative of the kinematic
sampling of the resonance in the integrations (e.g. see Dermer 1990;
Baring 1994).  For much of this spectral range, forward scattering in
the observer's frame is operating: \teq{\mu_f =\cos\Theta_f\approx 1}. 
This establishes almost Thomson kinematics, with the scattered photon
energy in the ERF satisfying \teq{\omega_f\approx\omega_i}. 
Substituting these approximations into Eqs.~(\ref{eq:sigma_pol})
and~(\ref{eq:TparTperp}), and then summing over final polarizations
yields a total approximate form for the flat portions of the spectral
production rate in Eq.~(\ref{eq:scatt_spec}) for cases \teq{\gamma_e\gg 1}:
\begin{equation}
   \dover{dn_{\gamma}}{dt\, d\erg_f} \;\approx\;  \dover{n_e n_s\, \sigt c}{\mu_+-\mu_-}
   \; \dover{3\pi B^2}{4\Gammacyc\,\gamma_e^3\erg_s^2}\quad .
 \label{eq:flattop_approx}
\end{equation}
The magnetic field and \teq{\gamma_e} dependences are evident in
Fig.~\ref{fig:upscatt_spectra} (remembering that the curves are
multiplied by \teq{\gamma_e^2} for the purposes of illustration), and
since \teq{\Gammacyc\propto B^2} when \teq{B\ll 1}, the normalization of
the flat spectrum is independent of the field strength in the magnetic
Thomson regime.  Only at the highest energies does the spectrum begin to
deviate from flat (i.e. horizontal) behavior, and this domain
corresponds to significant scattering angles in the ERF, i.e. cosines
\teq{1-\cos\theta_f} not much less than unity. Then the mathematical
form of the differential cross section becomes influential in
determining the spectral shape.  Specifically, for \teq{\theta_f \sim
\pi/2}, \teq{T_{\parallel}} drops far below \teq{T_{\perp}}, as is
evident in Eq.~(\ref{eq:TparTperp}), causing the observed dip in the
spectra. At slightly higher energies \teq{\erg_f}, there is a recovery
when \teq{T_{\parallel}} rises as \teq{\theta_f\to \pi}.  Observe that
\teq{T_\perp} is far less sensitive to scattering angles \teq{\theta_f}
in the ERF, except for supercritical fields when recoil becomes
significant. The sum of the two contributions yields a slight cusp at
the maximum energy \teq{\erg_f\approx \gamma_e(1+\beta_e)B/(1+2 B)} when
\teq{B \lesssim 0.5}, which disappears for higher fields when the recoil
reductions of \teq{d\sigma /d\cos\theta_f} become dominant.

In AXPs, the \teq{B=0.3} case best represents higher altitude locales for
the resonasphere, such as at smaller colatitudes near the polar axis. 
For \teq{B=3}, more typical of equatorial resonance locales, the flat
spectrum still appears at energies \teq{\erg_s\leq\erg_f \ll
\gamma_e(1+\beta_e)\, B/(1+2B)},  when again \teq{\cos\theta_f\approx
1}.  Yet the curves in Fig.~\ref{fig:upscatt_spectra} display more
prominent reductions at the uppermost energies \teq{\erg_f\sim
\gamma_e(1+\beta_e)\, B/(1+2 B)} due to the sampling of \teq{1/2 \leq
\omega_f \ll \omega_i} values in the ERF that correspond to strong
electron recoil effects.  Photons emitted in this regime have
\teq{1-\cos\theta_f\sim 1} in the ERF and are highly beamed along the
field in the observer's frame, as will become apparent shortly. At these
maximum energies, the approximations \teq{\cos\theta_f\approx 1} and
\teq{\omega_f\approx B/(1+2 B)} yield an analytic result for the
emission rate, precisely Eq.~(\ref{eq:flattop_approx}) multiplied by the
factor \teq{1/(1+2 B)^2} that controls the severity of the reduction at
the uppermost resonant energies. Note that Klein-Nishina reductions in
the cross section regulate the overall normalizations of the \teq{B=3}
curves, as does the fact that the cyclotron width \teq{\Gammacyc} no
longer scales with field strength as \teq{B^2}.

The dimensionless electron energy loss rate \teq{d\gamma_e/dt} can be
obtained by multiplying the differential spectrum \teq{dn_{\gamma}/dt
d\erg_f} in Eq.~(\ref{eq:scatt_spec}) by \teq{\erg_f/n_e} and
integrating over \teq{\erg_f}. Because of the flat nature of the
spectrum, this receives a dominant contribution from near the maximum
upscattered energy \teq{\gamma_e(1+\beta_e)\, B/(1+2 B)}.  Then, using
Eq.~(\ref{eq:flattop_approx}), in the magnetic Thomson limit \teq{B\ll
1}, since energies \teq{\erg_f\sim 2\gamma_eB} contribute most, the
result \teq{d\gamma_e/dt\propto B^4/\Gammacyc\propto B^2} is obviously
realized for \teq{\Gammacyc\approx 4\fsc B^2/3}.  The full integration
of Eq.~(\ref{eq:scatt_spec}) over \teq{\erg_f} (or equivalently
\teq{\cos\theta_f}) is analytically tractable, leading to
\begin{equation}
   \dover{d\gamma_e}{dt} \;\approx\;  \dover{n_s\, \sigt c}{\mu_+-\mu_-}
   \; \dover{3\pi B^2}{4\gamma_e\erg_s^2}\quad , \quad B\; \ll\; 1\quad .
 \label{eq:cool_rate}
\end{equation}
This result is commensurate with the form derived in Eq. (24) of Dermer 
(1990).  In contrast, when \teq{B\gg 1}, the overall normalization
of the spectrum at energies \teq{\erg_f\ll \gamma_e(1+\beta_e) B/(1+2 B)}
is still controlled by Eq.~(\ref{eq:flattop_approx}), but now the cyclotron 
decay rate dependence \teq{\Gammacyc\propto B^{1/2}} and recoil
reductions at the highest \teq{\erg_f} come into play, so that the cooling rate
\teq{d\gamma_e/dt} possesses a dependence on the field strength that
is much weaker than \teq{B^{3/2}}.

\begin{figure}
   \centering
   \includegraphics[width=0.47\textwidth]{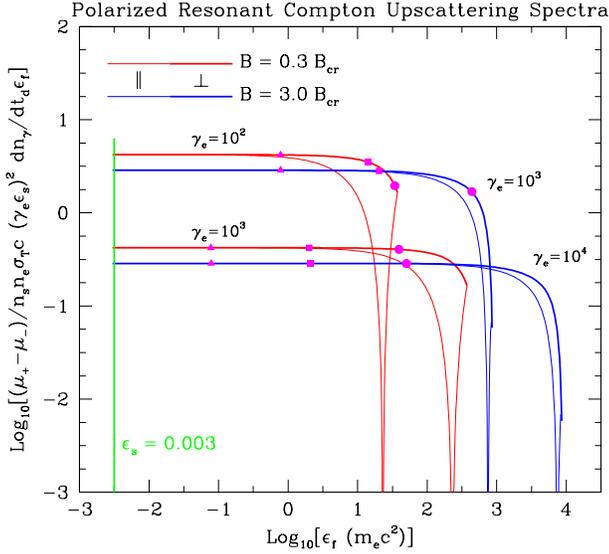}
   \vspace{-8pt}
   \caption{Resonant Compton upscattering spectra appropriate for an 
   AXP magnetosphere, scaled as in Fig.~\ref{fig:upscatt_spectra},
   but this time displaying the two polarizations \teq{\perp} (heavyweight) and
   \teq{\parallel} (lightweight) for the produced photons; the dominance of the
   \teq{\perp} polarization near the uppermost energies is evident.  Again, 
   the emergent photon energy \teq{\erg_f} is scaled in terms of \teq{m_ec^2}, 
   and results are presented for different relativistic electron Lorentz
   factors \teq{\gamma_e}, as labelled.  Specific emergent angles of the 
   emission in the observer's frame, with respect to the magnetic field
   direction, are indicated by the filled magenta symbols, with triangles denoting
   \teq{\Theta_f=5^{\circ}}, squares corresponding to \teq{\Theta_f=1^{\circ}},
   and circles representing \teq{\Theta_f=0.2^{\circ}}.  The magnetic field strengths 
   of \teq{B=3 B_{\rm cr}} (blue) and \teq{B=0.3 B_{\rm cr}} (red) correspond 
   perhaps to lower and higher altitudes, respectively.  Again, the green 
   vertical line marks the soft photon energy \teq{\erg_s=0.003}, with
   downscattering resonant emission at \teq{\erg_f < \erg_s} not being exhibited. 
   }
 \label{fig:upscatt_spectra_pol} 
\end{figure}

The resonant upscattering spectra are potentially polarized, perhaps
strongly.  Isolating the specific polarization forms in
Eq.~(\ref{eq:TparTperp}), the polarization-dependent resonant Compton
spectra are readily computed and are illustrated in
Fig.~\ref{fig:upscatt_spectra_pol}.  It is clear from
Eq.~(\ref{eq:TparTperp}) that the emissivities for photons of final
polarization state \teq{\perp} should always be superior to those for
the \teq{\parallel} state.  In a classical description, this is a
consequence of the physical ease with which an oscillating electron can
resonantly drive emission with electric field vectors perpendicular to
{\bf B}.  Yet for much of the range of emergent photon energies above
\teq{\erg_s=0.003}, there is no material difference between fluxes for
the two final polarizations.  This case corresponds to
\teq{\cos\theta_f\approx 1}, i.e. photon emission along the field in the
ERF, and hence induces zero linear polarization by symmetry, but permits
emergent circular polarization. At the highest produced energies,
significant differences between \teq{\perp} and \teq{\parallel} emission
appear, with \teq{1-\cos\theta_f} no longer very small (see
Eq.~(\ref{eq:TparTperp}) to help identify this characteristic). This
domain motivates the development of medium energy gamma-ray polarimeters
as a tool for geometry diagnostics.

Another feature of the upscattering process that is highlighted in the
Fig.~\ref{fig:upscatt_spectra_pol} is the intense beaming of radiation
along the field in the observer's frame of reference, and the profound
correlation of the angle of emission \teq{\Theta_f} with the emergent
photon energy \teq{\erg_f}.   This tight correspondence is
mathematically guaranteed by the appearance of the kinematic delta
function in Eq.~(\ref{eq:scatt_spec}) together with the intrinsically
narrow nature of the scattering resonance, which permits the delta
function approximation in Eq.~(\ref{eq:resonance_approx}).  The
extremely narrow range of \teq{\Theta_f} for each observed energy
\teq{\erg_f} is broadened when non-resonant scattering is introduced. 
In the resonant case here, the \teq{\gamma_e\gg 1} regime dictates that
most of the emission is collimated to within \teq{5^{\circ}} of the
field direction, and rapidly becomes beamed to within \teq{0.2^{\circ}}
as the final photon energy increases towards its maximum. This kinematic
characteristic guarantees that spectral formation in Compton
upscattering models is extremely sensitive to the observer's viewing
perspective in relation to the magnetospheric geometry, offering useful
probes if pulse-phase spectroscopy is achievable.

It should be noted that the spectra in Figs.~\ref{fig:upscatt_spectra}
and~\ref{fig:upscatt_spectra_pol} are in principal subject to
attenuation by magnetic pair production, \teq{\gamma\to e^+e^-},
reprocessing the highest energy photons to lower energies.  This process
is sensitive to the angle \teq{\Theta_f} of the scattered photons to the
field, which, from the results depicted here, is strongly coupled to
their energy \teq{\erg_f}.  Gonthier et al. (2000) demonstrated (see
their Figure~7) that generally, for an extended range of values for
\teq{B}, pair creation attenuation would only operate for
\teq{\omega_i\gtrsim 10}--30 in the ERF, since the resonant Compton
process couples the values of \teq{\omega_f} and \teq{\theta_f}. For the
resonant scattering considerations here, this criterion translates to
\teq{\gamma\to e^+e^-} being rife for local fields \teq{B\gtrsim 10} and
being marginal, or more probably ineffective, at lower field strengths.
For \teq{B\gtrsim 10} scattering circumstances, any pair creation that
ensues does so by generating pairs in the ground state (e.g. Usov \&
Melrose 1995; Baring \& Harding 2001), with Lorentz factors less than
\teq{\gamma_e} since \teq{\erg_f < \gamma_e}.  Accordingly the cascading
would consist at first of generations of upscattering and subsequent
pair creation, until high enough altitudes are encountered for pair
production to access excited Landau levels for the pairs, and then
synchrotron/cyclotron radiation can ensue and complicate the cascade.
\section{Discussion}
 \label{sec:discussion}

It is clear that the spectra exhibited in Fig.~\ref{fig:upscatt_spectra}
are considerably flatter than the hard X-ray tails (\teq{\sim
\erg_f^{-1}}) seen in the AXPs, and extend to energies much higher than
can be permitted by the Comptel upper bounds to these sources.  However,
they represent a preliminary indication of how flat the resonant
scattering process can render the emergent spectrum, and what an
observer detects will depend critically on his/her orientation and the
magnetospheric locale of the scattering. The one-to-one kinematic
correspondence between \teq{\erg_f} and \teq{\mu_f} (illustrated via the
filled symbols in Fig.~\ref{fig:upscatt_spectra_pol}), imposed by
\teq{\omega_f=\gamma_e \erg_f (1-\mu_f) = \omega'(\omega_i)} with
\teq{\omega_i=B}, implies that the highest energy photons are beamed
strongly along the local field direction.  This may or may not be
sampled by an instantaneous observation, which varies with the
rotational phase. Realistically, depending on the pulse phase, angles
corresponding to \teq{\mu_f <1} will be predominant, lowering the value
of \teq{\erg_f}.  Yet how low is presently unclear, and remains to be
explored via a model with full magnetospheric geometry, an essential
step.  One can also expect substantial spectral differences between
scattering locales attached to open and closed field lines, and also
between dipolar and more complicated field morphologies with smaller
radii of curvature.

Distributing the electron \teq{\gamma_e} such as through resonant
cooling will generate a convolution of the spectra depicted in
Fig.~\ref{fig:upscatt_spectra}; observe that the \teq{\gamma_e^2}
scaling of the y-axis implies that the normalization of the curves is a
strongly-declining function of \teq{\gamma_e}.  This can clearly steepen
the continuum for a particular range of \teq{\mu_f}.  For example, 
since \teq{\erg_f\propto \gamma_e} near the maximum photon energy 
for a fixed electron Lorentz factor, integration over a truncated 
\teq{\gamma_e^{-p}} distribution, where \teq{\gamma_e\geq \gamma_{e,min}},
naturally yields a photon spectrum \teq{\erg_f^{-(p+2)}} at energies
above the critical value \teq{\erg_f\sim 2\gamma_{e,min} B/(1+2 B)}
where the resonant flat top turns over.  In particular, cooling
the electrons as they propagate in the magnetosphere can
lead to significant and possibly dominant contributions from Lorentz
factors \teq{\gamma_e\lesssim 10} at higher altitudes and lower \teq{B}
that can evade the Comptel constraints on the AXPs (see Kuiper et al.
2006).  In the Thomson regime, the cooling tends to steepen the
continuum (e.g. Baring 1994) in the X-ray band due to a pile-up of
electrons at low \teq{\gamma_e}.  If electrons propagate to high
altitudes in magnetars, a similar steepening should be expected.  As the
polarized signal appears at the highest energies for each
\teq{\gamma_e}, a somewhat broad range of energies will exhibit
polarization above around 50--100 keV when integrating over an entire
cooled electron distribution.  Note also that the cooling may persist down
to mildly-relativistic energies, i.e., \teq{\gamma_e\sim 1}, in which
case it can seed a multiple scattering Comptonization of thermal X-rays 
that may generate the steep non-thermal continuum observed in AXPs 
below 10 keV. Such resonant, magnetic Comptonization has been 
explored by Lyutikov \& Gavriil (2006), and provides very good fits to 
both Chandra (Lyutikov \& Gavriil 2006) and XMM (see Rea et al. 2006) 
spectral data for the AXP 1E 1048.1-5937.

Finally, the introduction of non-resonant contributions can have a
significant impact on the spectral shape.  The resonasphere is spatially
confined, almost to a surface for a given photon trajectory, so that for
most of an X-ray photon's passage from the stellar surface, it scatters
only out of the cyclotron resonance with outward-going electrons,
particularly if rapid cooling is operating (a common circumstance).
Moreover, inward-moving electrons traversing closed field lines
participate in head-on collisions with surface X-rays, and so have great
difficulty accessing the resonance. These circumstances provide ample
opportunity for the emergent spectrum to develop a significant
non-resonant component that does not acquire the characteristically flat
spectral profiles exhibited here. Since Eq.~(\ref{eq:kinematic_range})
establishes \teq{\erg_f\lesssim 2\gamma_eB} for \teq{B\ll 1}, then this
upper bound becomes inferior to the classical, non-magnetic inverse
Compton result \teq{\erg_f\sim 4\gamma_e^2 \erg_s/3} when \teq{B
\lesssim \gamma_e\erg_s}.  This then defines a global criterion for when
non-resonant Compton cooling dominates the resonant process. Assessing
the relative weight of the resonant and non-resonant contributions
requires a detailed model of magnetospheric photon and electron
propagation and Compton scattering; this will form the focal point of
our upcoming modeling of the high energy emission tails from Anomalous
X-ray Pulsars.

\begin{acknowledgements}
	We thank Lucien Kuiper and Wim Hermsen for discussions concerning
	the INTEGRAL/RXTE data on the hard X-ray tails in Anomalous X-ray
	Pulsars, and the anonymous referee for suggestions helpful to the
	polishing of the manuscript.
\end{acknowledgements}



\begin{thebibliography}{3}
%

\def\aas{{Astron. Astrophys.}}
\def\aassupp{{Astron. Astrophys. Supp.}}
\def\apss{{Astr. Space Sci.}}
\def\apj{ApJ}
\def\nat{Nature}
\def\aaps{{Astron. \& Astr. Supp.}}
\def\apj{{ApJ}}
\def\aa{{A\&A}}
\def\apjs{{ApJS}}
\def\sp{{Solar Phys.}}
\def\jgr{{J. Geophys. Res.}}
\def\grl{{Geophys. Res. Lett.}}
\def\jphysb{{J. Phys. B}}
\def\ssr{{Space Science Rev.}}
\def\araa{{Ann. Rev. Astron. Astrophys.}}
\def\asr{{Adv. Space. Res.}}
\def\prc{{Phys. Rev. C}}
\def\prd{{Phys. Rev. D}}
\def\pr{{Phys. Rev.}}
\def\mnras{MNRAS}


\bibitem{baring94}
Baring, M.~G. 1994, in {\it Gamma-Ray Bursts}, eds. Fishman, G.,
	Hurley, K. \& Brainerd, J.~J., (AIP Conf. Proc. 307, New York) p. 572.

\bibitem{bgh05}
Baring, M.~G., Gonthier, P.~L., Harding A.~K. 2005, \apj,\vol{630}{430}

\bibitem{bh01}
Baring, M.~G. \& Harding A.~K. 2001, \apj,\vol{547}{929}

\bibitem{bs96}
Baykal, A. \& Swank, J. 1996, \apj,\vol{460}{470}

\bibitem{bs76}
Blandford, R.~D. \& Scharlemann, E.~T. 1976, \mnras\vol{174}{59}

\bibitem{bam86}
Bussard, R.~W., Alexander, S.~B. \& M\'esz\'aros, P 1986, \prd,\vol{34}{440}

\bibitem{clr71}
Canuto, V., Lodenquai, J. \& Ruderman, M., 1971, \prd,\vol{3}{2303}

\bibitem{dh86}
Daugherty, J.~K., \& Harding, A.~K. 1986, \apj,\vol{309}{362}

\bibitem{dh89}
Daugherty, J.~K., \& Harding, A.~K. 1989, \apj,\vol{336}{861}

\bibitem{derm90}
Dermer, C. D. 1990, \apj,\vol{360}{197}

\bibitem{dt92} 
Duncan, R. C. \& Thompson, C. 1992, \apj,\vol{392}{L9}

\bibitem{dr00}
Dyks, J. \& B. Rudak 2000, \aas\vol{360}{263}

\bibitem{gk04}
Gavriil, F.~P. \& Kaspi,ÊV.~M. 2004, \apj,\vol{609}{L67}

\bibitem{gkw02} 
Gavriil, F.~P., Kaspi,ÊV.~M. \& Woods, P.~M. 2002, \nat,\vol{419}{142}

\bibitem{gkw04} 
Gavriil, F.~P., Kaspi,ÊV.~M. \& Woods, P.~M. 2004, \apj,\vol{607}{959}

\bibitem{gj69} 
Goldreich, P. \& Julian, W. H. 1969, \apj,\vol{157}{869}

\bibitem{gonthier00} 
Gonthier, P.~L., Harding A.~K., Baring, M. G., et al. 
   2000, \apj,\vol{540}{907}
 %Costello, R. M. \& Mercer, C. L. 

\bibitem{hd06}
Harding, A.~K. \& Lai, D. 2006, Rep. Prog. Phys., \vol{69}{2631}

\bibitem{hm98}
Harding, A.~K. \& A.~G. Muslimov 1998,  \apj,\vol{508}{328}

\bibitem{herold79}
Herold, H. 1979, \prd,\vol{19}{2868}

\bibitem{hh05} 
Heyl, J. \& Hernquist, L.~E. 2005, \mnras,\vol{362}{777}

\bibitem{he89}
Ho, C, \& Epstein, R.~I. 1989, \apj,\vol{343}{227}

\bibitem{juett02} 
Juett, A.~M., Marshall, H.~L., Chakrabarty, D., et al. 2002, \apj,\vol{568}{L31}

\bibitem{kaspi03} 
Kaspi, V.~M., Gavriil, F.~P., Woods, P.~M., et al. 2003, \apj,\vol{588}{L93}

\bibitem{kuip06} 
Kuiper,ÊL., Hermsen,ÊW., den Hartog, P.~R., et al. 2006, \apj,\vol{645}{556}

\bibitem{kuip04} 
Kuiper,ÊL., Hermsen,ÊW. \& Mende\'z, M. 2004, \apj,\vol{613}{1173}

\bibitem{kulk03} 
Kulkarni, S.~R., Kaplan, D.~L., Marshall, H.~L., et al. 2003, \apj,\vol{585}{948}

\bibitem{latal86}
Latal, H.~G. 1986, \apj,\vol{309}{372}

\bibitem{lg06}
Lyutikov, M. \& Gavriil, F.~P. 2006, \mnras,\vol{368}{690}

\bibitem{mereg05} 
Mereghetti, S., G\"otz, D., Mirabel, I.~F., et al. 2005, \aa\  Lett., \vol{433}{L9}

\bibitem{molk05} 
Molkov, S., Hurley, K., Sunyaev, R., et al. 2005, \aa\  Lett., \vol{433}{L13}

\bibitem{ooster98}
Oosterbroek, T., Parmar, A.~N., Mereghetti, S., et al. 1998, \aas,\vol{334}{925}

\bibitem{patel03} 
Patel, S.~K., Kouveliotou, C.,  Woods, P.~M., et al. 2003, \apj,\vol{587}{367}

\bibitem{perna01} 
Perna, R., Heyl, J.~S., Hernquist, L.~E., et al. 2001, \apj,\vol{557}{18}

\bibitem{rea05}
Rea,ÊN., Oosterbroek,ÊT., Zane,ÊS., et al. 2005, \mnras, \vol{361}{710}

\bibitem{rea06}
Rea, N., Zane, S., Lyutikov, M., et al. 2006, \apss, in press. [{\tt astro-ph/0608650}]

\bibitem{sturn95}
Sturner, S.~J. 1995, \apj\vol{446}{292}
 
\bibitem{sdm95}
Sturner, S. J.,  Dermer, C. D. \& Michel, F. C. 1995, \apj,\vol{445}{736}

\bibitem{tb05} 
Thompson, C. \& Beloborodov, A.~M. 2005, \apj,\vol{634}{565}

\bibitem{td95} 
Thompson, C. \& Duncan, R. C. 1995, \mnras,\vol{275}{255}

\bibitem{td96} 
Thompson, C. \& Duncan, R. C. 1996, \apj,\vol{473}{332}

\bibitem{tiengo02} 
Tiengo, A., et al. 2002, \aa,\vol{383}{182}

\bibitem{um95}
Usov, V.~V. \& Melrose, D.~B. 1995, Aust. J. Phys., \vol{48}{571}

\bibitem{vg97} 
Vasisht, G. \& Gotthelf, E. V. 1997, \apj,\vol{486}{L129}


\end{thebibliography}
\end{document}